\documentclass[11pt,a4paper]{article}

\usepackage{times}
\usepackage{amsmath}
\usepackage{amssymb}
\usepackage{euler}
\begin{document}
\newcommand{\Cstar}{\mathfrak{C}}
\newcommand{\Hilbert}{\mathcal{H}}
\newcommand{\Obs}{\mathfrak{N}}            
\newcommand{\ObsComm}{\mathfrak{A}}        
\newcommand{\SubsetAlg}{\mathcal{S}}       
\newcommand{\CentreObs}{\mathfrak{N}^c} 
\newcommand{\Bounded}{\mathcal{B}}      
\newcommand{\Trace}{\mathrm{Tr}}        
\newcommand{\UnitComplex}{\mathrm{U}(1)}  

\title{Superselection Rules}
\author{Domenico Giulini         \\
University of Hannover           \\
Institute for Theoretical Physics\\
Appelstrasse 2, D-30167 Hannover, Germany\\
and\\
University of Bremen\\
ZARM\\
Am Fallturm 1, D-28359 Bremen, Germany}

\date{}

\maketitle

\begin{abstract}
\noindent
This note provides a summary of the meaning of the term 
`Superselection Rule' in Quantum Mechanics and Quantum-Field Theory. 
It is a slightly extended version of a contribution to the 
\emph{Compendium of Quantum Physics: Concepts, 
Experiments, History and Philosophy}, edited by Friedel Weinert, 
Klaus Hentschel, Daniel Greenberger, and Brigitte Falkenburg.
\end{abstract}

\subsection*{General Notion}
The notion of \textbf{superselection rule} (henceforth 
abbreviated SSR) was introduced in 1952 by Wick (1909-1992), 
Wightman, and Wigner (1902-1995)~\cite{Wick.Wightman.Wigner:1952} 
in connection with the problem of consistently assigning 
intrinsic parity to elementary particles. They understood 
an SSR as generally expressing ``restrictions on the nature and 
scope of possible measurements''. 

The concept of SSR should be contrasted with that of an ordinary 
 \textbf{selection rule}~(SR). The latter refers to a dynamical 
inhibition of some transition, usually due to the existence 
of a conserved quantity. Well known SRs in Quantum Mechanics
concern radiative transitions of atoms. For example, in case of electric 
dipole radiation they take the form $\Delta J=0,\pm1$ (except $J=0\rightarrow J=0$) 
and $\Delta M_J=0,\pm 1$. It says that the quantum numbers $J,M_J$ 
associated with the atom's total angular momentum may at most change 
by one unit. But this is only true for electric dipole transitions,
which, if allowed, represent the leading-order contribution in an 
approximation for wavelengths much larger than the size of the atom. 
The next-to-leading-order contributions are given by magnetic dipole
and electric quadrupole transitions, and for the latter $\Delta J=\pm 2$ 
\emph{is} possible. This is a typical situation as regards SRs: They 
are valid for the leading-order modes of transition, but not necessarily 
for higher order ones. 
In contrast, a SSR is usually thought of as making 
a more rigorous statement. It not only forbids certain transitions through 
particular modes, but altogether as a matter of some deeper lying 
principle; hence the ``Super''. In other words, transitions are not only 
inhibited for the particular dynamical evolution at hand, generated 
by the given Hamiltonian operator, but for all conceivable 
dynamical evolutions. 

More precisely, two states $\psi_1$ 
and $\psi_2$ are separated by a SR if $\langle\psi_1\mid H\mid\psi\rangle=0$ 
for the given Hamiltonian $H$. In case of the SR mentioned above, 
$H$ only contains the leading-order interaction between the radiation 
field and the atom, which is the electric dipole interaction. 
In contrast, the states are said to be separated by a SSR if 
$\langle\psi_1\mid A\mid\psi_2\rangle=0$ for \emph{all} 
(physically realisable) observables~$A$. This means that 
the relative phase between $\psi_1$ and $\psi_2$ is not measurable and 
that coherent superpositions of $\psi_1$ and $\psi_2$ cannot be 
verified or prepared. It should be noted that such a statement implies that the 
set of (physically realisable) observables is strictly smaller than 
the set of all self-adjoint operators on Hilbert 
space. For example,    
$A=\mid\psi_1\rangle\langle\psi_2\mid+\mid\psi_2\rangle\langle\psi_1\mid$
is clearly self-adjoint and satisfies 
$\langle\psi_1\mid A\mid\psi_2\rangle\ne 0$. 
Hence the statement of a SSR always implies
a restriction of the set of observables as compared to the set of all (bounded) 
self-adjoint operators on Hilbert space. In some sense, 
the existence of SSRs can be 
formulated in terms of observables alone (see below). 

Since all theories work with idealisations, the issue may be raised as 
to whether the distinction between SR and SSR is really well founded, 
or whether it could, after all, be understood as a matter of degree only. 
For example, dynamical decoherence is known to provide a 
very efficient mechanism for generating apparent SSRs, without assuming 
their existence on a fundamental 
level~\cite{Zurek:2003}\cite{Joos-etal:2003}.

\goodbreak
 
\subsection*{Elementary Theory}
\label{sec:Elementary Theory}
In the most simple case of only two \textbf{superselection sectors},
a SSR can be characterised by saying that the Hilbert space 
$\Hilbert$ decomposes as a direct sum of two orthogonal subspaces,
$\Hilbert=\Hilbert_1\oplus\Hilbert_2$, such that under the action 
of each observable vectors in $\Hilbert_{1,2}$ are 
transformed into vectors in $\Hilbert_{1,2}$ respectively. 
In other words, the action of observables in Hilbert space is 
reducible, which implies that 
$\langle\psi_1\mid A\mid\psi_2\rangle=0$ for each 
$\psi_{1,2}\in\Hilbert_{1,2}$ and all observables $A$. This 
constitutes an inhibition to the superposition principle
in the following sense: Let $\psi_{1,2}$ be normed vectors and 
$\psi_+=(\psi_1+\psi_2)/\sqrt{2}$, then  
\begin{equation}
\label{eq:InhibitSuperpos1}
 \langle\psi_+\mid A\mid\psi_+\rangle
=\tfrac{1}{2}\bigl(\langle\psi_1\mid A\mid\psi_1\rangle
 +\langle\psi_2\mid A\mid\psi_2\rangle\bigr)
=\Trace(\rho A)\,,
\end{equation}
where
\begin{equation}
\label{eq:InhibitSuperpos2}
\rho=\tfrac{1}{2}\bigl(\mid\psi_1\rangle\langle\psi_1\mid
    +\mid\psi_2\rangle\langle\psi_2\mid\bigr)\,.
\end{equation}
Hence, considered as state (expectation-value functional) on the 
given set of observables, the density matrix $\rho$ corresponding to 
$\psi_+$ can be written as non-trivial convex combination of the 
(pure) density matrices for $\psi_1$ and $\psi_2$ and therefore 
defines a mixed state rather than a pure state. 
Relative to the given 
observables, coherent superpositions of states in $\Hilbert_1$
with states in $\Hilbert_2$ do not exist. 

In direct generalisation, a characterisation of \textbf{discrete SSRs} 
can be given as follows: There exists a finite or countably infinite 
family $\{P_i\mid i\in I\}$ of mutually orthogonal ($P_iP_j=0$ 
for $i\ne j$) and exhaustive ($\sum_{i\in I}P_i=1$) 
projection operators ($P_i^\dagger=P_i$, $P_i^2=P_i$) 
on Hilbert space $\Hilbert$, such that each observable 
commutes with all $P_i$. Equivalently, one may also say 
that states on the given set of observables (here represented 
by density matrices) commute with all $P_i$, which is 
equivalent to the identity   
\begin{equation}
\label{eq:SSR-EquivStatements-a}
\rho=\sum_i P_i\rho P_i\,.
\end{equation}
We define $\lambda_i:=\Trace(\rho P_i)$ and let $I'\subset I$ 
be the subset of indices $i$ for which $\lambda_i\ne 0$. 
If we further set $\rho_i:=P_i\rho P_i/\lambda_i$ for $i\in I'$, 
then (\ref{eq:SSR-EquivStatements-a}) is equivalent to 
\begin{equation}
\label{eq:ReducedDensityMatrix}
\rho=\sum_{i\in I'}\lambda_i\rho_i\,,
\end{equation}
showing that $\rho$ is a non-trivial convex combination 
if $I'$ contains more than one element. The only pure 
states are the projectors onto rays within a single 
$\Hilbert_i$. 
In other words, only vectors (or rays) in the union (not the 
linear span) $\bigcup_{i\in I}\Hilbert_i$ can correspond 
to pure states. If, conversely, \emph{any} non-zero 
vector in this union defines a pure state, with different 
rays corresponding to different states, one speaks of an 
\textbf{abelian superselection rule}. The $\Hilbert_i$ 
are then called \textbf{superselection sectors} or 
\textbf{coherent subspaces} on which the observables act 
irreducibly. The subset $Z$ of observables commuting with 
all observables is then given by 
$Z:=\bigl\{\sum_ia_iP_i\mid a_i\in\mathbb{R}\bigr\}$.
They are called \textbf{superselection-} or 
\textbf{classical observables}. 

In the general case of \textbf{continuous SSRs} $\Hilbert$ 
splits as direct integral of an uncountable set of Hilbert 
spaces $\Hilbert(\lambda)$, where $\lambda$ is an element 
of some measure space $\Lambda$, so that 
\begin{equation}
\label{eq:HilbertDirectIntegral}
\Hilbert=\int_\Lambda d\mu(\lambda)\Hilbert(\lambda)\,,
\end{equation}
with some measure $d\mu$ on $\Lambda$. Observables 
are functions $\lambda\mapsto O(\lambda)$, with $O(\lambda)$ 
acting on $\Hilbert(\lambda)$. Closed subspaces of 
$\Hilbert$ left invariant by the observables are precisely 
given by
\begin{equation}
\label{eq:ContinuousSSR-Subspaces}
\Hilbert(\Delta)=\int_{\Delta}
d\mu(\lambda)\Hilbert(\lambda)\,,
\end{equation}
where $\Delta\subset\Lambda$ is any measurable 
subset of non-zero measure. In general, a single
$\Hilbert(\lambda)$ will not be a subspace (unless
the measure has discrete support at $\lambda$).

In the literature, SSRs are discussed in connection with
a variety of superselection-observables, most notably univalence, 
overall mass (in non-relativistic QM), electric charge, baryonic 
and leptonic charge, and also time. 

\subsection*{Algebraic Theory}
\label{sec:AlgebraicTheory}
In Algebraic Quantum Mechanics, a system 
is characterised by a $C^*$--algebra $\Cstar$. Depending on 
contextual physical conditions, one chooses a faithful 
representation $\pi:\Cstar\rightarrow\Bounded(\Hilbert)$ 
in the (von Neumann) algebra of bounded operators on Hilbert 
space $\Hilbert$. After completing the image of 
$\pi$ in the weak operator-topology on $\Bounded(\Hilbert)$
(a procedure sometimes called \textbf{dressing} of 
$\Cstar$~\cite{Bogolubov.etal:GPoQFT}) one obtains a von 
Neumann sub-algebra $\Obs\subset\Bounded(\Hilbert)$, 
called the \textbf{algebra of (bounded) observables}. 
The physical observables proper correspond to the 
self-adjoint elements of $\Obs$. 

The \textbf{commutant} $\SubsetAlg'$ of any subset 
$\SubsetAlg\subseteq\Bounded(\Hilbert)$ is defined by 
\begin{equation}
\label{eq:Def:CommutatntOfSet}
\SubsetAlg':=\{A\in\Bounded(\Hilbert)\mid AB=BA\,,
\forall B\in\SubsetAlg\}\,,
\end{equation}
which is automatically a von Neumann algebra. One calls 
$\SubsetAlg'':=(\SubsetAlg')'$ the von Neumann algebra generated 
by $\SubsetAlg$. It is the smallest von Neumann sub-algebra
of $\Bounded(\Hilbert)$ containing $\SubsetAlg$, so that if 
$\SubsetAlg$ was already a von Neumann algebra one has 
$\SubsetAlg''=\SubsetAlg$; in particular, $\bigl(\pi(\Cstar)\bigr)''=\Obs$

SSRs are now said to exists iff\footnote{Throughout we use `iff' as abbreviation
for `if and only if'.} the commutant $\Obs'$ is not trivial, 
that is, iff $\Obs'$ is different from multiples of the unit operator. 
Projectors in $\Obs'$ then define the sectors. 
\textbf{Abelian SSRs} are characterised by $\Obs'$ being abelian. The significance of 
this will be explained below. $\Obs'$ is often referred to as 
\textbf{gauge algebra}. Sometimes the algebra of physical observables 
is \emph{defined} as the commutant of a given gauge algebra. That the 
gauge algebra is abelian is equivalent to $\Obs'\subseteq\Obs''=\Obs$ so 
that $\Obs'=\Obs\cap\Obs'=:\CentreObs$, the centre
of $\Obs$. An abelian $\Obs'$ is equivalent to \textbf{Dirac's requirement}, 
that there should exist a complete set of commuting 
observables~\cite{Jauch:1960} (cf. Chap.\,6 of~\cite{Joos-etal:2003}). 

In finite-dimensional Hilbert spaces Dirac's requirement is equivalent 
to the hypothesis that there be sufficiently many pairwise commuting 
self-adjoint elements of $\Obs$ so that the simultaneous eigenspaces
are one-dimensional. In other words, each array of eigenvalues 
(`quantum numbers'), one for each self-adjoint element, uniquely 
determines a pure quantum state (a ray in $\Hilbert$). This implies 
the existence of a self-adjoint $A\in\Obs$ with a simple spectrum
(pairwise distinct eigenvalues). It then follows that any other 
self-adjoint $B\in\Obs$ commuting with $A$ must then be a 
polynomial function (of degree $n-1$ if $n=\mathrm{dim}(\Hilbert)$) 
of $A$ and that there exists a vector $\psi\in\Hilbert$ so that any 
other $\phi\in\Hilbert$ is obtained by applying a polynomial 
(of degree $n-1$) in $A$ to $\psi$. The vector $\psi$ is called a 
cyclic vector for $\Obs$ and may be chosen to be any vector with 
non-vanishing components in each simultaneous eigenspace for the 
complete set of commuting observables; see 
Chap.\,6 of~\cite{Joos-etal:2003} for details. 
 
The algebraic theory allows to translate these statements to the 
general situation. Here, the existence of a `complete' set of 
commuting observables is interpreted as existence of a `maximal' 
abelian subalgebra $\ObsComm\subset\Obs$. Here it is crucial that 
`maximal' is properly understood, namely as `maximal in                  
$\Bounded(\Hilbert)$' and not just maximal in $\Obs$, which 
would be a rather trivial requirement (given Zorn's lemma, a
maximal abelian subalgebra in $\Obs$ always exists). Now, it is 
easy to see that $\ObsComm$ is maximal abelian (in 
$\Bounded(\Hilbert)$) iff it is equal to its commutant 
(in $\Bounded(\Hilbert)$):
\begin{equation}
\label{eq:CondMaxAbelian1}
\text{$\ObsComm$ max. abelian}\,\Leftrightarrow\,
\ObsComm=\ObsComm'\,.
\end{equation}
This is true since $\ObsComm\subseteq\ObsComm'$ certainly holds
due to $\ObsComm$ being abelian. On the other hand, 
$\ObsComm\supseteq\ObsComm'$ also holds since it just expresses 
the maximality requirement that $\ObsComm$ already contains all 
elements of $\Bounded(\Hilbert)$ commuting with each element of  
$\ObsComm$. 

Moreover, it can be shown that the existence of a maximal abelian 
subalgebra $\ObsComm$ in $\Obs$ is equivalent to $\Obs'$ being 
abelian: 
\begin{equation}
\label{eq:CondMaxAbelian2}
\text{$\ObsComm$ max. abelian}\,\Leftrightarrow\,
\Obs'\subseteq\Obs''=\Obs\,.
\end{equation}
The proof of this important statement is easy enough to be reproduced 
here: Suppose first that $\ObsComm=\ObsComm'$, then
$\Obs\supseteq\ObsComm=\ObsComm'\supseteq\Obs'$ and hence 
$\Obs'\subseteq\Obs=\Obs''$, implying that $\Obs'$ is abelian. 
Conversely, suppose $\Obs'$ is abelian: 
\begin{equation}
\label{eq:ObsAbelian}
\Obs'\subseteq\Obs
\qquad\text{($\Obs'$ is abelian)}\,.
\end{equation}
Choose an abelian subalgebra $\ObsComm\subseteq\Obs$ which is 
maximal in $\Obs$:
\begin{equation}
\label{eq:ObsCommMaxAbelian}
\ObsComm=\ObsComm'\cap\Obs
\qquad\text{($\ObsComm$ max. abelian in $\Obs$)}\,.
\end{equation}
As already noted above, Zorn's lemma guarantees the existence 
of $\ObsComm$ satisfying~(\ref{eq:ObsCommMaxAbelian}). We show that 
 $\ObsComm$, albeit only required to be maximal in $\Obs$, is in 
fact maximal in $\Bounded(\Hilbert)$ due to $\Obs'$ being abelian. 
Indeed, since $\ObsComm\subseteq\Obs$ trivially implies 
$\Obs'\subseteq\ObsComm'$, we have    
\begin{equation}
\Obs'\ \mathop{=}^{(\ref{eq:ObsAbelian})}\ \Obs\cap\Obs'
\subseteq\Obs\cap\ObsComm'\ \mathop{=}^{(\ref{eq:ObsCommMaxAbelian})}\
\ObsComm\,.
\end{equation}
Since $\Obs'\subseteq\ObsComm$ trivially implies 
$\ObsComm'\subseteq\Obs$, equation (\ref{eq:ObsCommMaxAbelian}) 
immediately leads to $\ObsComm=\ObsComm'$.
This shows that Dirac's requirement is equivalent to the 
hypothesis of abelian SSRs. 

Another requirement equivalent to Dirac's is that there 
should exist a \textbf{cyclic vector} $\psi\in\Hilbert$ for 
$\Obs$. This means that the smallest closed subspace of 
$\Hilbert$ containing $\Obs\psi:=\{A\psi\mid A\in\Obs\}$ 
is $\Hilbert$ itself. 

In Quantum Logic a quantum system is characterised 
by the lattice of propositions (corresponding to the closed 
subspaces, or the associated projectors, in Hilbert-space 
language). The subset of all propositions which are 
compatible with all other propositions is called the 
\textbf{centre of the lattice}. It forms a Boolean sub-lattice. 
A lattice is called \textbf{irreducible} iff its centre is trivial 
(i.e. just consists of $0$, the smallest lattice element). The 
presence of SSRs is now characterised by a non-trivial centre. 
Propositions in the centre are sometimes called \textbf{classical}.

\subsection*{SSRs and Conserved Additive Quantities}
\label{sec:SSR-CAQ}
Let $Q$ be the operator of some charge-like quantity that 
behaves additively under composition of systems and also 
shares the property that the charge of one subsystem is 
independent of the state of the complementary subsystem (here we 
restrict attention to two subsystems). This implies that if 
$\Hilbert=\Hilbert_1\otimes\Hilbert$ is the Hilbert space 
of the total system and $\Hilbert_{1,2}$ those of the 
subsystems, $Q$ must be of the form $Q=Q_1\otimes 1+1\otimes Q_2$, 
where $Q_{1,2}$ are the charge operators of the subsystems. 
We also assume $Q$ to be conserved, i.e. to commute with 
the total Hamiltonian that generates time evolution on 
$\Hilbert$. It is then easy to show that a SSR for $Q$ 
persists under 
the operations of composition, decomposition, and time 
evolution: If the density matrices $\rho_{1,2}$ commute 
with $Q_{1,2}$ respectively, then, trivially, 
$\rho=\rho_1\otimes\rho_2$ commutes with $Q$. Likewise, 
if $\rho$ (not necessarily of the form $\rho_1\otimes\rho_2$) 
commutes with $Q$, then the reduced density matrices 
$\rho_{1,2}:=\Trace_{2,1}(\rho)$ (where $\Trace_i$ stands 
for tracing over $\Hilbert_i$) commute with $Q_{1,2}$
respectively. This shows that if states violating the SSR 
cannot be prepared initially (for whatever reason, 
not yet explained), they cannot be created though 
subsystem interactions~\cite{Wick.Wightman.Wigner:1970}.   
This has a direct relevance for measurement theory, 
since it is well known that an exact von~Neumann measurement 
of an observable $P_1$ in system\,1 by system\,2 is possible only 
if $P_1$ commutes with $Q_1$, and that an approximate measurement 
is possible only insofar as system\,2 can be prepared in a 
superposition of $Q_2$ eigenstates~\cite{Araki.Yanase:1960}.


Let us see how to prove the second to last statement for the 
case of discrete spectra. Let $S$ be the system to be measured, $A$ 
the measuring apparatus and $\Hilbert=\Hilbert_S\otimes\Hilbert_A$
the Hilbert space of the system plus apparatus. The charge-like 
quantity is represented by the operator $Q=Q_S\otimes 1+1\otimes Q_A$, 
the observable of $S$ by $P\in\Bounded(\Hilbert_S)$.
Let $\{\vert s_n\rangle\}\subset\Hilbert_S$ be a set of normalised 
eigenstates for $P$ so that $P\vert s_n\rangle=p_n\vert s_n\rangle$. Let 
$U\in\Bounded(\Hilbert)$ be the unitary evolution operator for the 
von Neumann measurement and $\{\vert a_n\rangle\}\subset\Hilbert_A$ a set of 
normalised `pointer states' with neutral pointer-position $a_0$, so that 
\begin{equation}
\label{eq:MeasurementEvolution}
U\bigl(\vert s_n\rangle\vert a_0\rangle\bigr)=
\vert s_n\rangle\vert a_n\rangle
\end{equation}
We assume the total $Q$ to be conserved during the measurement, i.e. 
$[U,Q]=0$. Clearly $\langle a_n\mid a_m\rangle\ne 1$ if $n\ne m$, for, otherwise, 
this process is not a measurement at all, since $\langle a_n\mid a_m\rangle=1$
iff $\vert a_n\rangle=\vert a_n\rangle$. Let now $n\ne m$, then the following 
lines prove the claim:
\begin{alignat}{1}
(p_n-p_m)\langle s_n\vert Q_S\vert s_m\rangle
&\,=\,(p_n-p_m)\langle s_n\vert\langle a_0\vert\, Q\,
                      \vert s_m\rangle\vert a_0\rangle\nonumber \\
&\,=\,(p_n-p_m)\langle s_n\vert\langle a_0\vert\ U^\dagger QU\
                      \vert s_m\rangle\vert a_0\rangle\nonumber \\
&\,=\,(p_n-p_m)\langle s_n\vert\langle a_n\vert\ Q_S\otimes 1+1\otimes Q_A\
  \vert s_m\rangle\vert a_m\rangle\qquad\nonumber \\
\label{eq:MeasurmentTheorem}
&\,=\,\langle a_n\vert a_m\rangle\,(p_n-p_m)
\langle s_n\vert Q_S\vert s_m\rangle\,.
\end{alignat}
The first and fourth equality follow from $\langle s_n\vert s_m\rangle =0$, 
the second from $[U,Q]=0$ and unitarity of $U$, and the third from 
(\ref{eq:MeasurementEvolution}). Equality of the left-hand side with the last 
expression on the right-hand side, taking into account 
$\langle a_n\vert a_m\rangle\ne 1$, is possible iff $p_n\ne p_m$ implies 
$\langle s_n\vert Q_S\vert s_m\rangle=0$, which means that $Q_s$ is reduced
by (i.e. acts within) each eigenspace of $P$, which in turn implies that 
$Q_s$ commutes with $P$, as was to be shown.  

As already indicated, the reasoning above does not explain 
the actual existence of SSRs in the presence of conserved additive 
quantities, for it does not imply anything about the \emph{initial} 
nonexistence of SSR violating states. In fact, there are many 
additive conserved quantities, like momentum and angular momentum, 
for which certainly no SSRs is at work. The crucial observation 
here is that the latter quantities are physically always 
understood as \emph{relative} to a system of reference that, 
by its very definition, must have certain localisation properties 
which exclude the total system to be in eigenstate of 
\emph{relative} (linear and angular) momenta. Similarly it 
was argued that one may have superpositions of relatively charged 
states~\cite{Aharonov.Susskind:1967a}. A more complete 
account of this conceptually important point, including a
comprehensive list of references, is given in 
Chap.\,6 of~\cite{Joos-etal:2003}.

\subsection*{SSRs and Symmetries}
\label{sec:SSR-Symmetries}
Symmetries in Quantum Mechanics are often implemented 
via \textbf{unitary ray-representations} rather than 
proper unitary representations (here we discard anti-unitary 
ray-representations for simplicity). A unitary 
ray-representation is a map $U$ from the symmetry group $G$ 
into the group of unitary operators on Hilbert space 
$\Hilbert$ such that the usual condition of homomorphy,  
$U(g_1)U(g_2)=U(g_1g_2)$, is generalised to 
\begin{equation}
\label{eq:RayRep1}
U(g_1)U(g_2)=\omega(g_1,g_2)\,U(g_1g_2)\,,
\end{equation}
where $\omega:G\times G\rightarrow\UnitComplex:=
\{z\in\mathbb{C}\mid \vert z\vert=1\}$ is the so-called
\textbf{multiplier} that satisfies 
\begin{equation}
\label{eq:RayRep2}
\omega(g_1,g_2)\omega(g_1g_2,g_3)=\omega(g_1,g_2g_3)\omega(g_2,g_3)\,,
\end{equation}
for all $g_1,g_2,g_3$ in $G$, so as to ensures associativity: 
$U(g_1)\bigl(U(g_2)U(g_3)\bigr)=\bigl(U(g_1)U(g_2)\bigr)U(g_3)$.
Any function $\alpha:G\rightarrow\UnitComplex$ allows to 
redefine $U\mapsto U'$ via $U'(g):=\alpha(g)U(g)$, 
which amounts to a redefinition $\omega\mapsto\omega'$
of multipliers given by   
\begin{equation}
\label{eq:RayRep3}
\omega'(g_1,g_2)
=\tfrac{\alpha(g_1)\alpha(g_2)}{\alpha(g_1g_2)}\
\omega(g_1,g_2)\,.
\end{equation}
Two multipliers $\omega$ and $\omega'$ are called \textbf{similar} 
iff \eqref{eq:RayRep3} holds for some function
$\alpha$. A multiplier is called \textbf{trivial} iff 
it is similar to $\omega\equiv 1$, in which case the 
ray-representation is, in fact, a proper representation in 
disguise.  

The following result is now easy to show: Given unitary 
ray-representations $U_{1,2}$ of $G$ on $\Hilbert_{1,2}$, 
respectively, with non-similar multipliers $\omega_{1,2}$,
then no ray-representation of $G$ on 
$\Hilbert=\Hilbert_1\oplus\Hilbert_2$ exists which 
restricts to $U_{1,2}$ on $\Hilbert_{1,2}$ respectively. 
From this a SSR follows from the requirement that the 
Hilbert space of pure states should carry a ray-representation 
of $G$, since such a space cannot contain invariant linear 
subspaces that carry ray-representations with non-similar 
multipliers.   

An example is given by the SSR of univalence, that 
is, between states of integer and half-integer spin. Here $G$ is 
the group $SO(3)$ of proper spatial rotations. For integer spin 
it is represented by proper unitary representations, for half 
integer spin with non-trivial multipliers. Another often quoted 
example is the Galilei group, which is implemented in 
non-relativistic quantum mechanics by non-trivial unitary 
ray-representations whose multipliers depend on the total mass 
of the system and are not similar for different masses. 

Such derivations have sometimes been criticised (e.g. in 
\cite{Weinberg:QToF1}) for depending crucially on ones 
prejudice of what the symmetry group $G$ should be. 
The relevant observation here is the following: Any 
ray-representation of $G$ can be made into a proper 
representation of a larger group $\bar G$, which is  
a \emph{central extension} of $G$. But no superselection rules 
follow if $\bar G$ rather than $G$ were required to be the 
acting symmetry group on the set of pure states. For example, 
in case of the rotation group, $G=SO(3)$, it is sufficient 
to take $\bar G=SU(2)$, its double (and universal) cover. 
For $G$ the 10-parameter inhomogeneous  Galilei group it is 
sufficient to take for $\bar G$ an extension by the additive 
group $\mathbb{R}$, which may even be motivated on classical 
grounds~\cite{Giulini:1996}.

\subsection*{SSRs in Local Quantum Field Theories}
\label{sec:SSR-Locality}
In Quantum Field Theory SSRs can 
arise from the restriction to (quasi) local observables. 
Charges which can be measured by fluxes through closed surfaces 
at arbitrarily large spatial distances must then commute 
with all observables. A typical example is given by the 
total electric charge, which is given by the integral over 
space of the local charge density $\rho$. According to 
Maxwell's equations, the latter equals the divergence of 
the electric field $\vec E$, so that Gau{\ss}' theorem 
allows to write
\begin{equation}
\label{eq:Q-SSR}
Q=\lim_{R\mapsto\infty}\int_{\Vert\vec x\Vert=R}
(\vec n\cdot\vec E)d\sigma\,,
\end{equation}
where $\vec n$ is the normal to the sphere $\Vert\vec x\Vert=R$
and $d\sigma$ its surface measure. If $A$ is a local observable its  
support is in the \emph{causal complement} of the spheres 
$\Vert\vec x\Vert=R$ for sufficiently large $R$. Hence, in the 
quantum theory, $A$ commutes with $Q$. It is possible, though 
technically far from trivial, that this formal reasoning can 
indeed be justified in Local Quantum Field 
Theory~\cite{Strocchi.Wightman:1974}. For example, one 
difficulty is that Gau{\ss}' law does not hold as an operator 
identity. 

In modern Local Quantum-Field Theory 
\cite{Haag:LocQuantPhys}, representations of the quasi-local 
algebra of observables are constructed through the choice of 
a preferred state on that algebra (GNS-construction), like 
the Poincar\'e invariant vacuum state, giving rise to the 
\textbf{vacuum sector}. The superselection structure is restricted 
by putting certain selection conditions on such states, like 
e.g. the Doplicher-Haag-Roberts (DHR) selection criterion
for theories with mass gap (there are various 
generalisations~\cite{Haag:LocQuantPhys}), according to which 
any representation should be unitarily equivalent to the 
vacuum representation when restricted to observables whose 
support lies in the causal complement of a sufficiently large 
(causally complete) bounded region in spacetime. Interestingly 
this can be closely related to the existence of gauge groups
whose equivalence classes of irreducible unitary representations 
faithfully label the superselection sectors. Recently, a 
systematic study of SSRs in `Locally Covariant 
Quantum Field Theory' was started in~\cite{Brunetti.Ruzzi:2007}.
Finally we mention that SSRs may also arise as a consequence of 
non-trivial spacetime topology~\cite{Ashtekar.Sen:1980}.

\subsection*{Environmentally Induced SSRs}
The ubiquitous mechanism of decoherence effectively 
restricts the \emph{local} verification of 
coherences~\cite{Joos-etal:2003}. For example, scattering of
light on a particle undergoing a two-slit experiment 
\emph{delocalises} the relative-phase information for the two 
beams along with the escaping light. Hence effective SSRs 
emerge locally in a practically irreversible manner, 
albeit the correlations are actually never destroyed but 
merely delocalised. The emergence of effective 
SSRs through the dynamical process of decoherence has also been 
called \textbf{einselection}~\cite{Zurek:2003}. For example, 
this idea has been applied to the problem of why certain molecules 
naturally occur in eigenstates of chirality rather than 
energy and parity, i.e. why sectors of different 
chirality seem to be superselected so that chirality becomes a 
classical observable. This is just a special case of the general 
question of how classical behaviour can emerge in 
Quantum Theory. It may be asked whether \emph{all} SSRs are eventually 
of this dynamically emergent nature, or whether strictly fundamental 
SSRs persist on a kinematical level~\cite{Joos-etal:2003}. 
The complementary situation in theoretic modelling may be 
characterised as follows: Derivations of SSRs from axiomatic 
formalisms lead to exact results on models of only approximate 
validity, whereas the dynamical approach leads to approximate 
results on more realistic models.

\subsection*{SSRs in Quantum Information}
In the theory of Quantum Information a somewhat softer 
variant of SSRs is defined to be a restriction on the allowed local 
operations (completely positive 
and trace-preserving maps on density matrices) on a 
system~\cite{Bartlett.Wiseman:2003}. In general, it therefore 
leads to constraints on (bipartite) entanglement. Here the 
restrictions considered are usually not thought of as being 
of any fundamental nature, but rather for mere practical reasons. 
For example, without an external reference system for the 
definition of an overall spatial orientation, only `rotationally 
covariant' operations $\mathcal{O}:\rho\mapsto\mathcal{O}(\rho)$
are allowed, which means that $\mathcal{O}$ must satisfy  
\begin{equation}
\label{eq:CovOperation}
\mathcal{O}\bigl[U(g)\rho U^\dagger(g)\bigr]
=U(g)\mathcal{O}(\rho)U^\dagger(g)\quad\forall g\in SO(3)\,,
\end{equation}
where $U$ is the unitary representation of the group 
$SO(3)$ of spatial rotations in Hilbert space. 
Insofar as the local situation is concerned, this may be 
rephrased in terms of the original setting of SSRs, e.g. by 
regarding $SO(3)$ as gauge group, restricting local 
observables and states to those commuting with $SO(3)$. 
On the other hand, one also wishes to consider situations in 
which, for example, a local bipartite system (Alice and Bob) 
is given a state that has been prepared by a third party 
that is \emph{not} subject to the SSR.         

\newpage
\small

\begin{thebibliography}{10}

\bibitem{Aharonov.Susskind:1967a}
Yakir Aharonov and Leonard Susskind.
\newblock Charge superselection rule.
\newblock {\em Physical Review}, 155(5):1428--1431, 1967.
%
\bibitem{Araki.Yanase:1960}
Huzihiro Araki and Yanase~Mutsuo M.
\newblock Measurement of quantum mechanical operators.
\newblock {\em Physical Review}, 120(2):622--626, 1960.
%
\bibitem{Ashtekar.Sen:1980}
Abhay Ashtekar and Amitabha Sen.
\newblock On the role of space-time topology in quantum phenomena:
  Superselection of charge and emergence of nontrivial vacua.
\newblock {\em Journal of Mathematical Physics}, 21(3):526--533, 1980.
%
\bibitem{Bartlett.Wiseman:2003}
Stephen~D. Bartlett and Howard~M. Wiseman.
\newblock Entanglement constrained by superselection rules.
\newblock {\em Physical Review Letters}, 91(9):097903, 2003.
%
\bibitem{Bogolubov.etal:GPoQFT}
Nikolai~Nikolaevich Bogolubov et~al.
\newblock {\em General Principles of Quantum Field Theory}, volume~10 of {\em
  Mathematical Physics and Applied Mathematics}.
\newblock Kluwer Academic Publishers, Dordrecht, 1990.
%
\bibitem{Brunetti.Ruzzi:2007}
Romeo Brunetti and Giuseppe Ruzzi.
\newblock Superselection sectors and general covariance.\,{I}.
\newblock {\em Communications in Mathematical Physics}, 270(1):69--108, 2007.
%
\bibitem{Giulini:1996}
Domenico Giulini.
\newblock On {Galilei} invariance in quantum mechanics and the {Bargmann}
  superselection rule.
\newblock {\em Annals of Physics (New York)}, 249(1):222--235, 1996.
%
\bibitem{Haag:LocQuantPhys}
Rudolf Haag.
\newblock {\em Local Quantum Physics: Fields, Particles, Algebras}.
\newblock Texts and Monographs in Physics. Springer Verlag, Berlin, second
  revised and enlarged edition, 1996.
%
\bibitem{Jauch:1960}
Josef~Maria Jauch.
\newblock Systems of observables in quantum mechanics.
\newblock {\em Helvetica Physica Acta}, 33:711--726, 1960.
%
\bibitem{Joos-etal:2003}
Erich Joos, {H.-Dieter} Zeh, Claus Kiefer, Domenico Giulini, Joachim Kupsch,
  and Ion-Olimpiu Stamatescu.
\newblock {\em Decoherence and the Appearence of a Classical World in Quantum
  Theory}.
\newblock Springer Verlag, Berlin, second edition, 2003.
%
\bibitem{Strocchi.Wightman:1974}
Franco Strocchi and Arthur~Strong Wightman.
\newblock Proof of the charge superselection rule in local relativistic quantum
  field theory.
\newblock {\em Journal of Mathematical Physics}, 15(12):2198--2224, 1974.
\newblock Erratum: ibid 17(10):1930-1931, 1976.
%
\bibitem{Weinberg:QToF1}
Steven Weinberg.
\newblock {\em The Quantum Theory of Fields. Volume\,I Foundations}.
\newblock Cambridge University Press, Cambridge, 1995.
%
\bibitem{Wick.Wightman.Wigner:1952}
Gian~Carlo Wick, Arthur~Strong Wightman, and Eugene~Paul Wigner.
\newblock The intrinsic parity of elementary particles.
\newblock {\em Physical Review}, 88(1):101--105, 1952.
%
\bibitem{Wick.Wightman.Wigner:1970}
Gian~Carlo Wick, Arthur~Strong Wightman, and Eugene~Paul Wigner.
\newblock Superselection rule for charge.
\newblock {\em Physical Review~D}, 1(12):3267--3269, 1970.
%
\bibitem{Zurek:2003}
Wojciech~Hubert Zurek.
\newblock Decoherence, einselection, and the quantum origins of the classical.
\newblock {\em Reviews of Modern Physics}, 75:715--775, 2003.
%
\end{thebibliography}


\end{document}